\documentclass[epj]{svjour}
\usepackage{latexsym}
\usepackage{graphics}
\usepackage{epsfig}
\usepackage{rotating}
\usepackage{longtable}

\begin{document}
\title{Role of cross-shell excitations in the reaction $^{54}$Fe(\overrightarrow{\rm d},p)$^{55}$Fe}
\author{M. Mahgoub$^1$ \and R. Kr\"ucken$^1$ \and Th. Faestermann$^1$ \and A.
Bergmaier$^2$ \and D. Bucurescu$^3$ \and R. Hertenberger$^4$ \and
Th. Kr\"oll$^1$ \and H.-F. Wirth$^{1,4}$  \and A. F.
Lisetskiy$^{5,6}$
}                     
%
%
\institute{Physik Department, Technische Universit\"at M\"unchen,
D-85748 Garching, Germany
 \and
Institut f\"ur Angewandte Physik und Messtechnik, Universit\"at der
Bundeswehr M\"unchen, D-85577 Neubiberg, Germany
 \and
Horia Hulubei National Institute of Physics and Nuclear Engineering
(IFIN-HH), R-77125 Bucharest, Romania
 \and
Department f\"ur Physik, Ludwig Maximilians Universit\"at M\"unchen,
D-85748 Garching, Germany
 \and
Gesellschaft f\"ur Schwerionenforschung, D-64291 Darmstadt, Germany
 \and
Department of Physics, University of Arizona, Tucson, AZ 85721, USA}

\date{Received: date / Revised version: date}
%
\abstract{ The reaction $^{54}{\rm Fe}(\overrightarrow{\rm d},{\rm
p})^{55}{\rm Fe}$ was studied at the Munich Q3D spectrograph with a
14 MeV polarized deuteron beam. Excitation energies, angular
distributions and analyzing powers were measured for 39 states up to
4.5 MeV excitation energy. Spin and parity assignments were made and
spectroscopic factors deduced by comparison to DWBA calculations.
The results were compared to predictions by large scale shell model
calculations in the full pf-shell and it was found that reasonable
agreement for energies and spectroscopic factors below 2.5 MeV could
only be obtained if up to 6 particles were allowed to be excited
from the f$_{7/2}$ orbital into p$_{3/2}$, f$_{5/2}$, and p$_{1/2}$
orbitals across the $N=28$ gap. For levels above 2.5 MeV the
experimental strength distribution was found to be significantly
more fragmented than predicted by the shell model calculations.
} 
\authorrunning{M. Mahgoub et al.}
\titlerunning{$^{54}$Fe(\overrightarrow{\rm d},p)$^{55}$Fe}
\maketitle

\section{Introduction}
In the simplest spherical shell model approach the $N=Z=28$ nucleus
$^{56}$Ni has the properties of a doubly-magic core. However,
evidence for the softness of the $^{56}$Ni core has been obtained
experimentally \cite{Kra94} and theoretically \cite{Ots98}. With the
availability of more computing power and the development of the new
effective interaction GXPF1 for the pf-shell \cite{Hon02} core
excitations were found to play a significant role in the structure
of nuclei in the vicinity of $^{56}$Ni\cite{Hon04}. Experimental
evidence for the crucial role of cross-shell excitations for the
yrast spectra was, for example, reported in $^{58}$Cu \cite{Lis03}.
The stability of the magic number 28 is also of astrophysical
importance, {\em e.g.} for the electron capture rates in supernova
explosions (see e.g. Ref. \cite{Lan03}).

The current paper reports on a precision study of the excited states
of $^{55}$Fe using the $^{54}{\rm Fe}(\overrightarrow{\rm d},{\rm
p})^{55}{\rm Fe}$ reaction. While this reaction has been performed
several times in the past \cite{Koc72,Jun08,Tay80}, those studies
had limited energy resolution and sensitivity and therefore,
spectroscopic factors and definite spin assignments were obtained
only for a limited number of excited states. Those previous results
were typically compared to shell model calculations that assumed a
good $^{56}$Ni core and thus the sensitivity to cross-shell
excitations was not tested.

The excellent energy resolution and sensitivity of the Munich Q3D
magnetic spectrograph used in this study enabled the discovery of
several new states and the determination of spins for a number of
known states. For many states spectroscopic factors were determined
for the first time. The data are compared to the results of large
scale shell model calculations using the GXPF1 effective
interaction. Reasonable agreement for energies and spectroscopic
factors was achieved for states up to about 2.5 MeV when at least
6-particle 6-hole (6p-6h) excitations across the N=28 shell gap were
taken into account, while the experimental results could not be
satisfactory reproduced if less than six particles are promoted
across the shell gap. Thus, the results of our study clearly
demonstrate the importance of $^{56}$Ni core excitations for
$^{54}$Fe and $^{55}$Fe.

\section{Experimental Details}

The $^{54}{\rm Fe}(\overrightarrow{\rm d},{\rm p})^{55}{\rm Fe}$
reaction was studied by bombarding a 100$\mu{\rm g/cm^2}$ thick
94.6\% isotopically enriched self supporting $^{54}$Fe target with
polarized deuterons from a Stern-Gerlach polarized ion source
\cite{Her05} and accelerated to 14 MeV by the
MLL\footnote{Maier-Leibnitz Laboratory of the Technische
Universit\"at M\"unchen and the Ludwig-Maximilians-Universit\"at
M\"unchen} MP-Tandem Van de Graaff accelerator. The reaction
products were analyzed with the Munich Q3D spectrograph \cite{Lof73}
and then detected in a 1\,m long cathode strip focal-plane detector
\cite{Wir00,Wir01} with $\Delta E$-$E_{rest}$ particle
identification and position determination. The acceptance solid
angle of the spectrograph was 11.7 msr (horizontaly 54 mrad), except
for the most forward angle (5$^{\circ}$) where it was 5.7 msr.
Typical beam currents were around 0.5 $\mu$A on target.

\begin{figure}[htbp]
\resizebox{0.5\textwidth}{!}{%
  \includegraphics{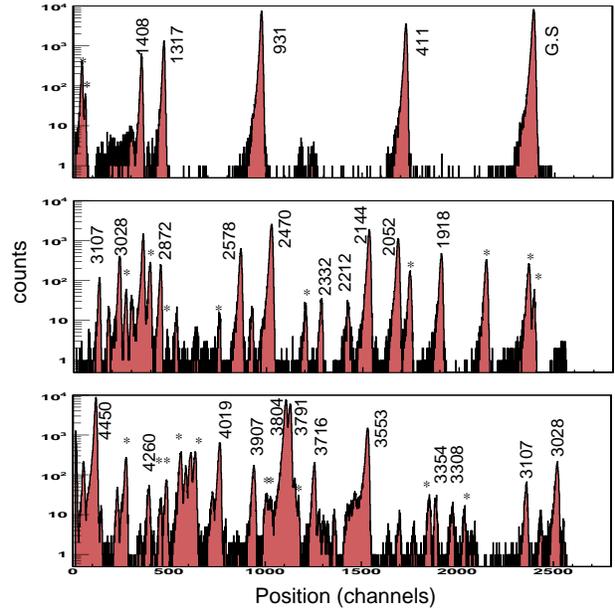}}
\caption{Focal plane spectra of the
Q3D observed at 30$^\circ$ with spin-up polarized deuteron beam for the reaction
$^{54}$Fe$( \protect\overrightarrow{\rm d} ,{\rm p} )^{55}$Fe. Magnetic field settings
were chosen such that excitation energies of
600 keV (top), 2200 keV (middle), and 3600 keV (bottom) were centered on the Q3D focal plane.
Energies for some levels, as obtained in this work, are given in keV. Lines marked with * are from the $^{56}$Fe impurities in the target.
\label{fig:fe55spec}
    }
\end{figure}

Spectra were measured at 8 angles between 5$^\circ$ and 40$^\circ$
in 5$^\circ$ steps plus one measurement at 50$^\circ$. For each
angle six spectra were collected, for 3 different magnetic settings,
covering  the excitation energy range from 0 to $\approx$4.5 MeV and
for spin-up and spin-down polarization of the deuteron beam,
respectively. Figure \ref{fig:fe55spec} shows Q3D spectra for those
settings taken at an angle of 30$^\circ$. The overall FWHM energy
resolution was around 7 keV, being mostly determined by the target
thickness. The spectra were essentially background free. All runs
were normalized to the beam current integrated by a Faraday cup
placed behind the target. The spectra were calibrated by using the
energies of those states in $^{55}$Fe that were known to better than
1 keV \cite{Jun08}. In the energy range from 3.0 to 4.5 MeV only
three levels with sufficient energy are known at 3072.0, 3108.7,
3552.3 keV. However, as will be discussed below, there are four
energy peaks previously assigned to $^{55}$Fe excitation energies at
3285, 3860, 4123, and 4372 keV which actually result from the
contaminant reaction $^{56}$Fe(d,p)$^{57}$Fe present in this energy
range, enabling an accurate energy calibration.

Differential reaction cross-sections for each angle were obtained by
averaging the cross-sections obtained for spin-up ($\frac{d \sigma}{
d \Omega} \uparrow$) or spin-down ($\frac{d \sigma}{ d \Omega}
\downarrow$) polarized beam. In addition to determine angular
distributions, the polarized beam with a polarization of $p=0.65$
enabled calculating the analyzing power at each angle using the
relation
\begin{equation}
A_{\rm y} = \frac{2}{3p} \cdot
\frac{ {\frac{d \sigma}{ d \Omega} \uparrow} - {{\frac{d \sigma}{ d \Omega} \downarrow}} }
{{\frac{d \sigma}{ d \Omega} \uparrow} + {{\frac{d \sigma}{ d \Omega} \downarrow}} }.
\label{eq:Ay}
\end{equation}

\section{Results}

In total 39 levels in $^{55}$Fe were observed in the energy range
from 0 to 4450 keV, of which 6 levels are observed for the first
time. For the other levels various amounts of information were
available ranging from only the energy, a tentative or firm spin
assignment to spectroscopic factors in some cases.

For five states that were previously assigned to $^{55}$Fe at 2015,
3285, 3860, 4123, and 4372 keV we could show that the observed lines
actually belong to known levels in $^{57}$Fe at energies of 366,
1627, 2220, 2456, and 2758 keV. The lines resulted from the 5.1 \%
$^{56}$Fe component in our target. We proved the assignment to
$^{57}$Fe by comparing the spectra for the $(d,p)$ reaction from our
94.6\% isotopically enriched $^{54}$Fe target with those obtained
with a 99.9\% isotopically enriched $^{56}$Fe target under exactly
same experimental conditions. The assignment of those levels to
$^{55}$Fe \cite{Jun08,Tay80,Mac60,Spe64} was based on transfer
experiments that used targets with less than 95\% $^{54}$Fe isotopic
enrichment and thus lines of $^{57}$Fe would be expected in the
spectra. However, we did not find any note in the original work that
a direct comparison was performed with a pure $^{56}$Fe target in
order to identify contributions from the $^{56}$Fe$(d,p)$ reaction
in the spectra. We are therefore confident that the levels at 2015,
3285, 3860, 4123, and 4372 keV do not exist in $^{55}$Fe.

\begin{figure}[htbp]
\resizebox{0.5\textwidth}{!}{%
 \centering
  \includegraphics{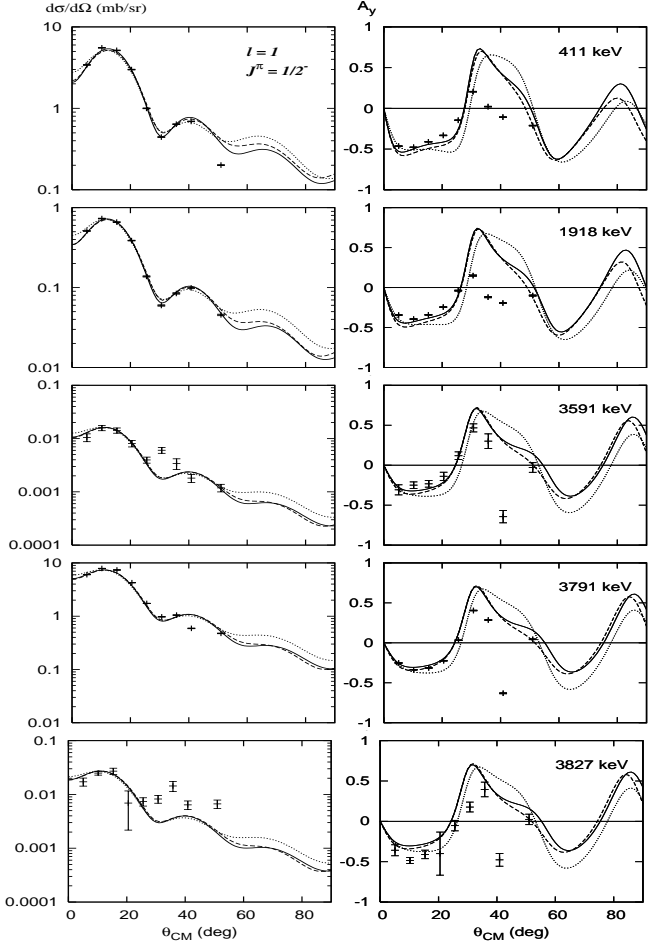}}
  \caption{Differential cross-section
  $\frac{d \sigma}{d \Omega}$ and the analyzing power $A_{\rm y}$ for
  $^{55}{\rm Fe}$ levels with $\Delta l=1$ and $J^\pi=1/2^-$.
  Curves indicate the DWBA calculations by CHUCK3 using the three different
  sets of OM-potentials from Refs.\cite{Per76} (full curve),
  \cite{Tay80} (dotted curve) and
  \cite{Fes92} (dashed curve).
    \label{fig:55Fe-L1}
    }
\end{figure}

\begin{figure}[htbp]
\resizebox{0.5\textwidth}{!}{%
 \centering
  \includegraphics{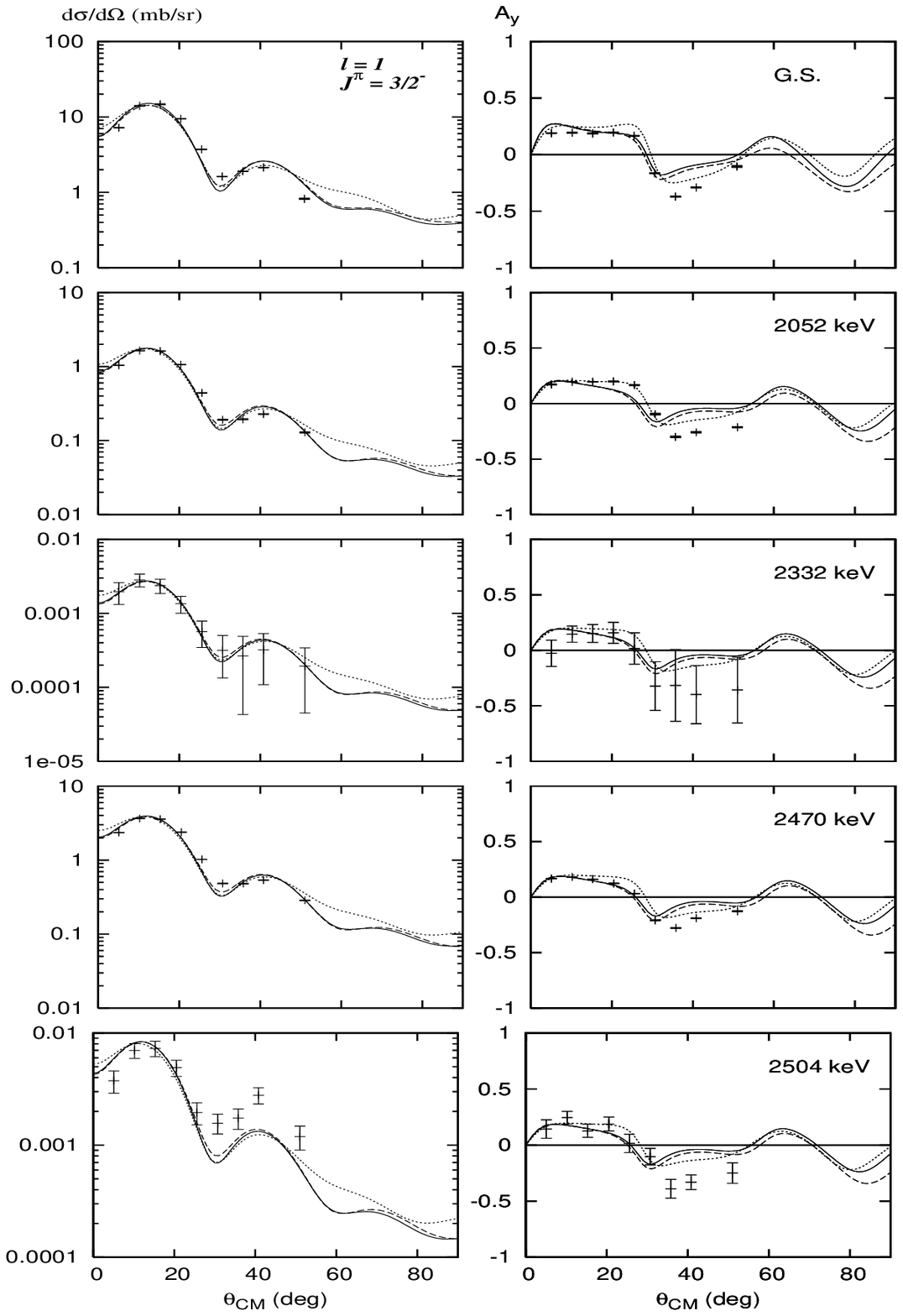}}
  \caption{Same as Fig. \ref{fig:55Fe-L1} for states with $\Delta l=1$ and $J^\pi=3/2^-$ and energies below 3.0 MeV.
    \label{fig:55Fe-L1J3a}
    }
\end{figure}

\begin{figure}[htbp]
\resizebox{0.5\textwidth}{!}{%
 \centering
  \includegraphics{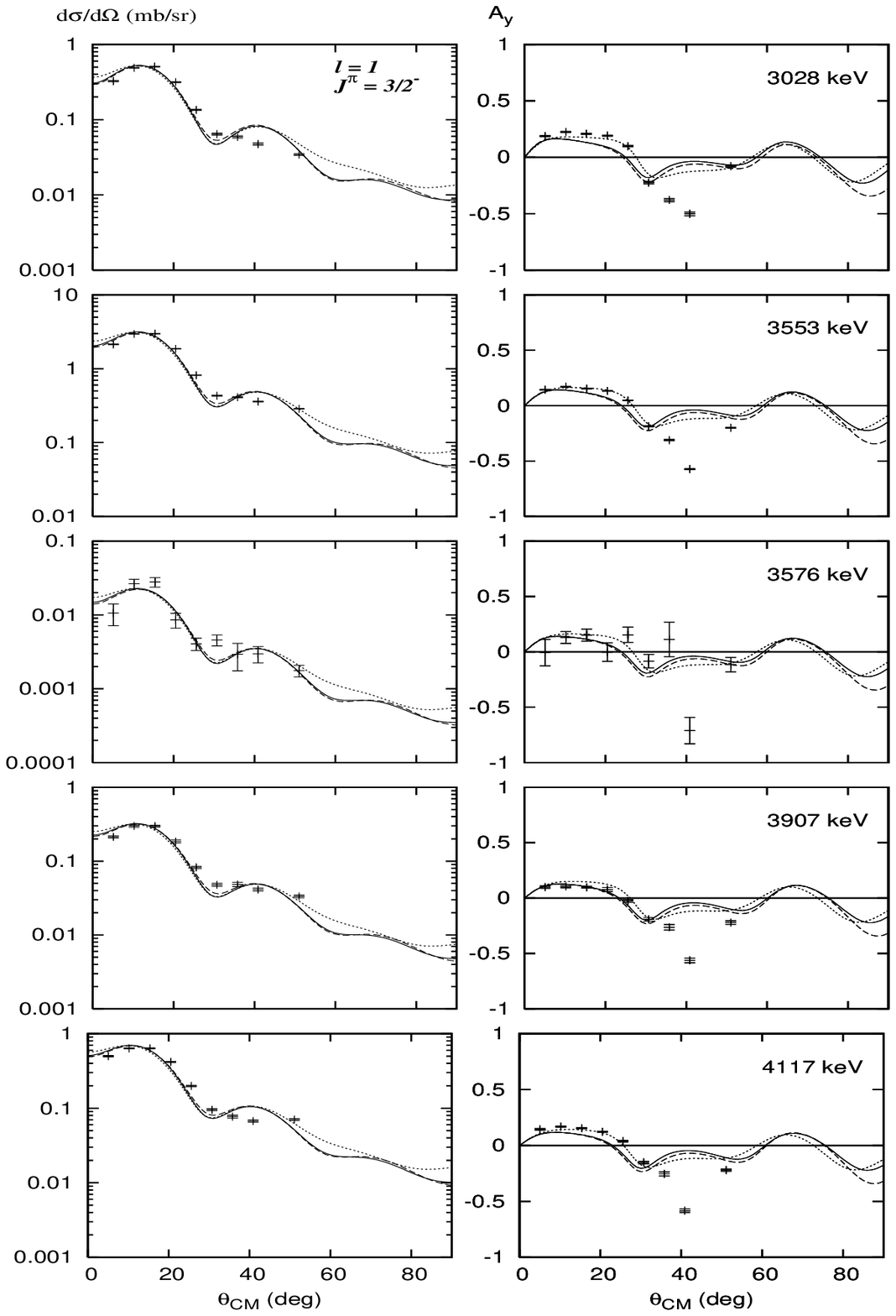}}
  \caption{Same as Fig. \ref{fig:55Fe-L1} for states with $\Delta l=1$ and  $J^\pi=3/2^-$ and energies above 3.0 MeV.
    \label{fig:55Fe-L1J3b}
    }
\end{figure}

\begin{figure}[htbp]
\resizebox{0.5\textwidth}{!}{%
 \centering
  \includegraphics{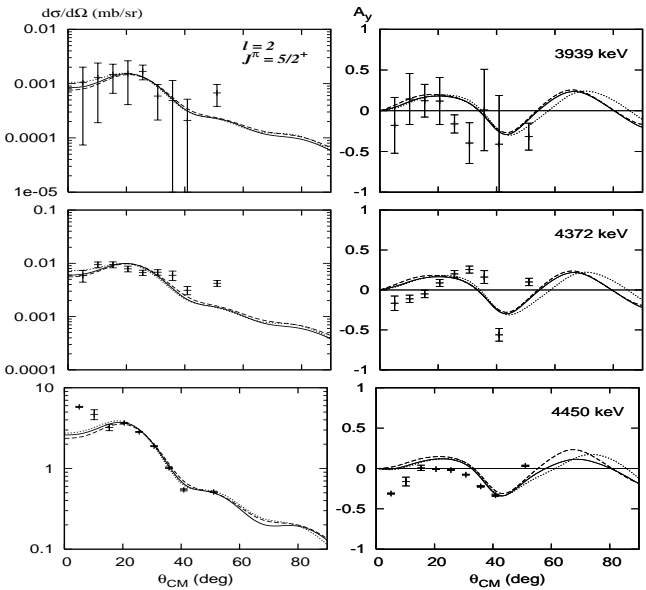}}
  \caption{Same as Fig. \ref{fig:55Fe-L1} for states with $\Delta l=2$ and $J^\pi=3/2^+$ and $J^\pi=5/2^+$.
    \label{fig:55Fe-L1J2}
    }
\end{figure}

\begin{figure}[htbp]
\resizebox{0.5\textwidth}{!}{%
 \centering
  \includegraphics{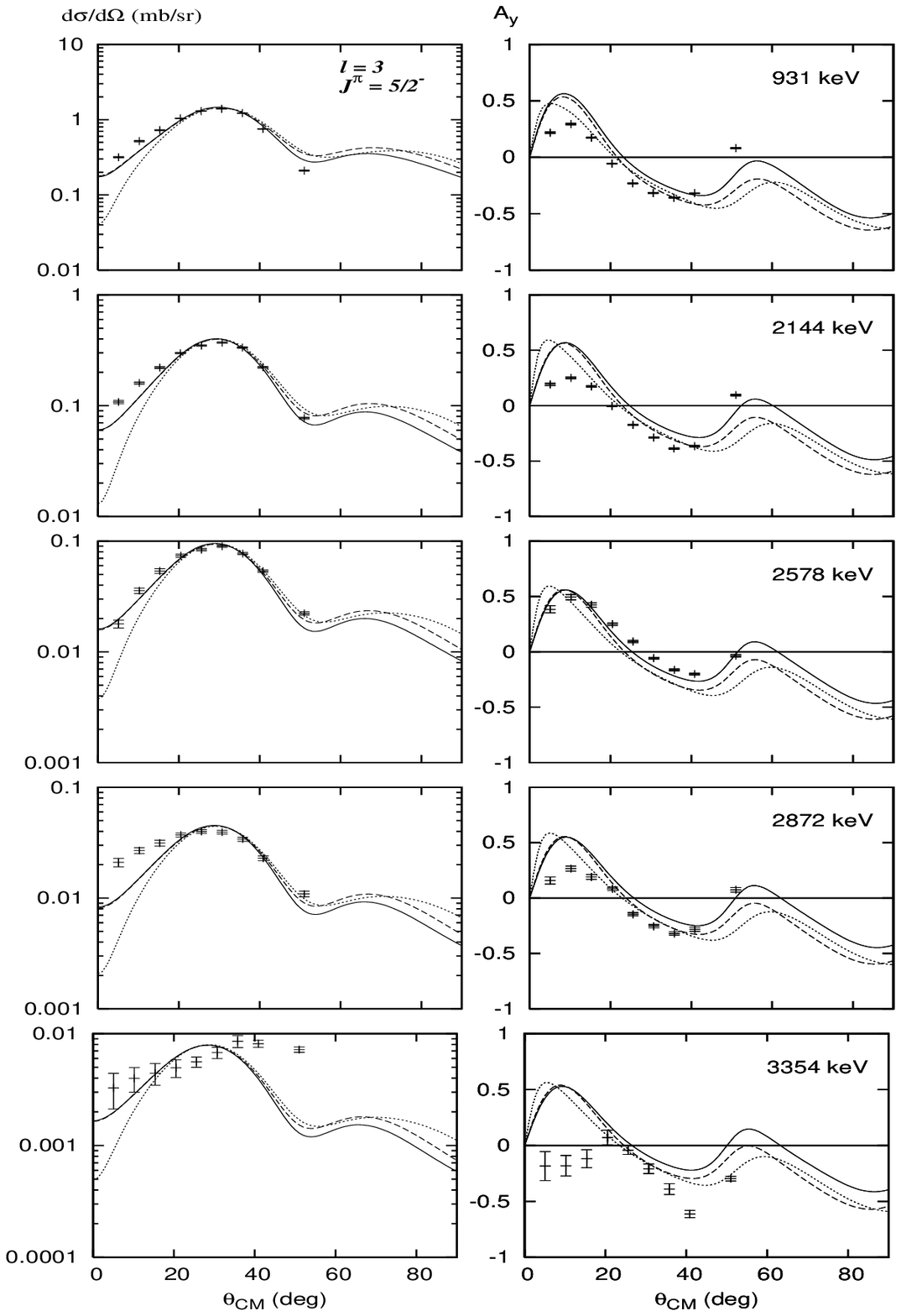}}
  \caption{Same as Fig. \ref{fig:55Fe-L1} for states with $\Delta l=3$ and  $J^\pi=5/2^-$ and energies below 3.7 MeV.
    \label{fig:55Fe-L3J5a}
    }
\end{figure}

\begin{figure}[htbp]
\resizebox{0.5\textwidth}{!}{%
 \centering
  \includegraphics{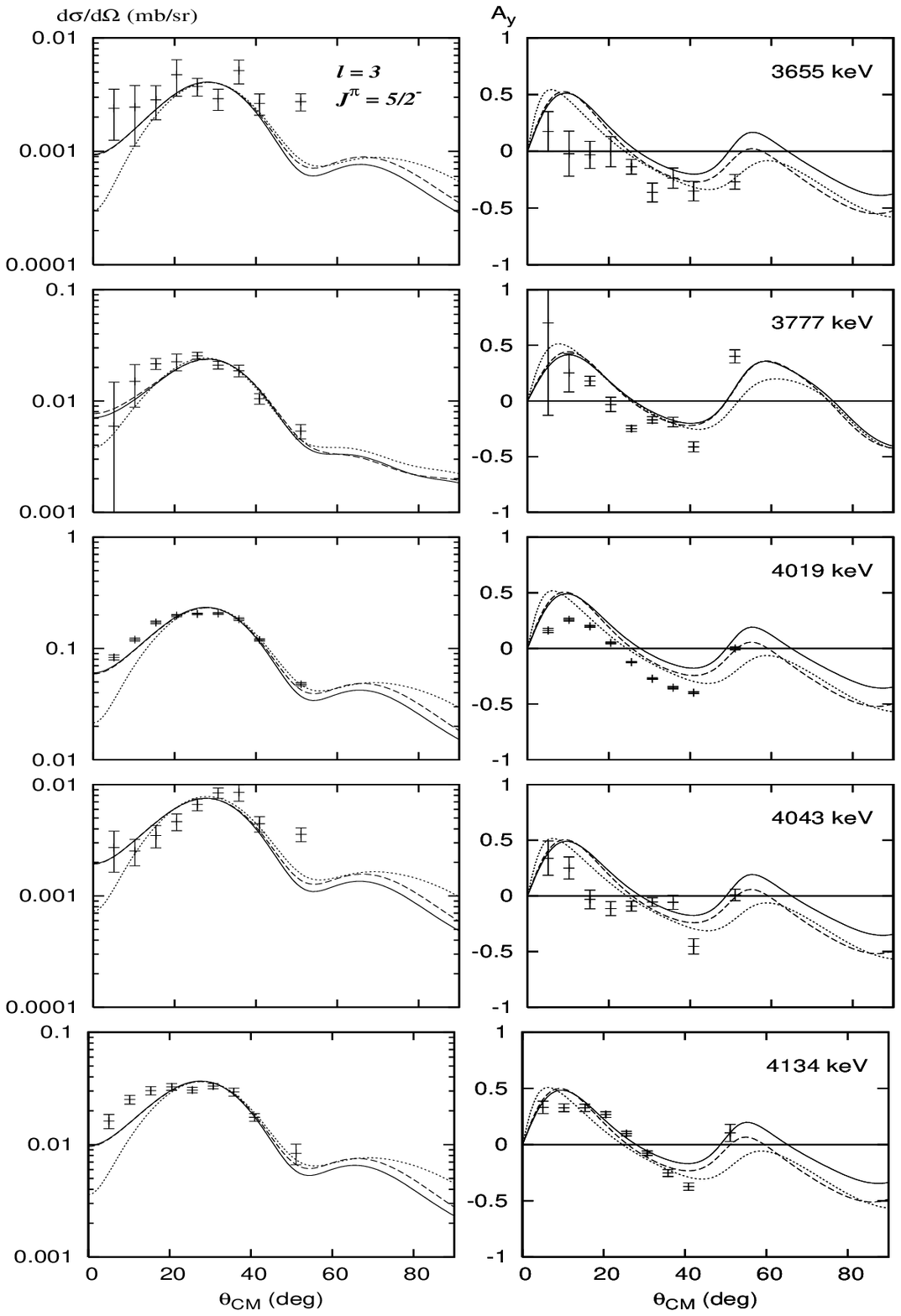}}
  \caption{Same as Fig. \ref{fig:55Fe-L1} for states with $\Delta l=3$ and  $J^\pi=5/2^-$ and energies above 3.7 MeV.
    \label{fig:55Fe-L3J5b}
    }
\end{figure}

\begin{figure}[htbp]
\resizebox{0.5\textwidth}{!}{%
 \centering
  \includegraphics{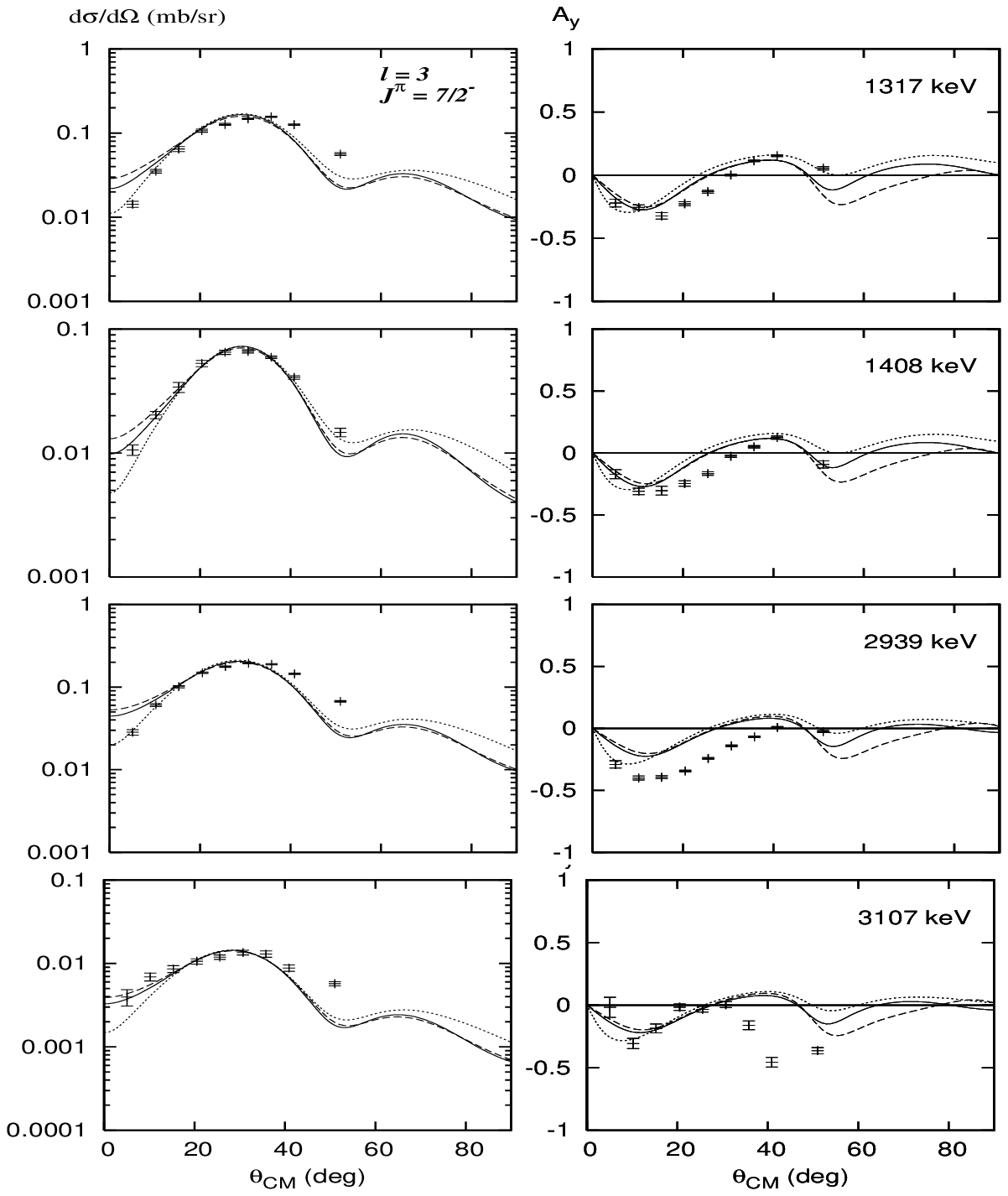}}
  \caption{Same as Fig. \ref{fig:55Fe-L1} for states with $\Delta l=3$ and  $J^\pi=7/2^-$ and energies below 3.2 MeV.
    \label{fig:55Fe-L3J7a}
    }
\end{figure}

\begin{figure}[htbp]
\resizebox{0.5\textwidth}{!}{%
 \centering
  \includegraphics{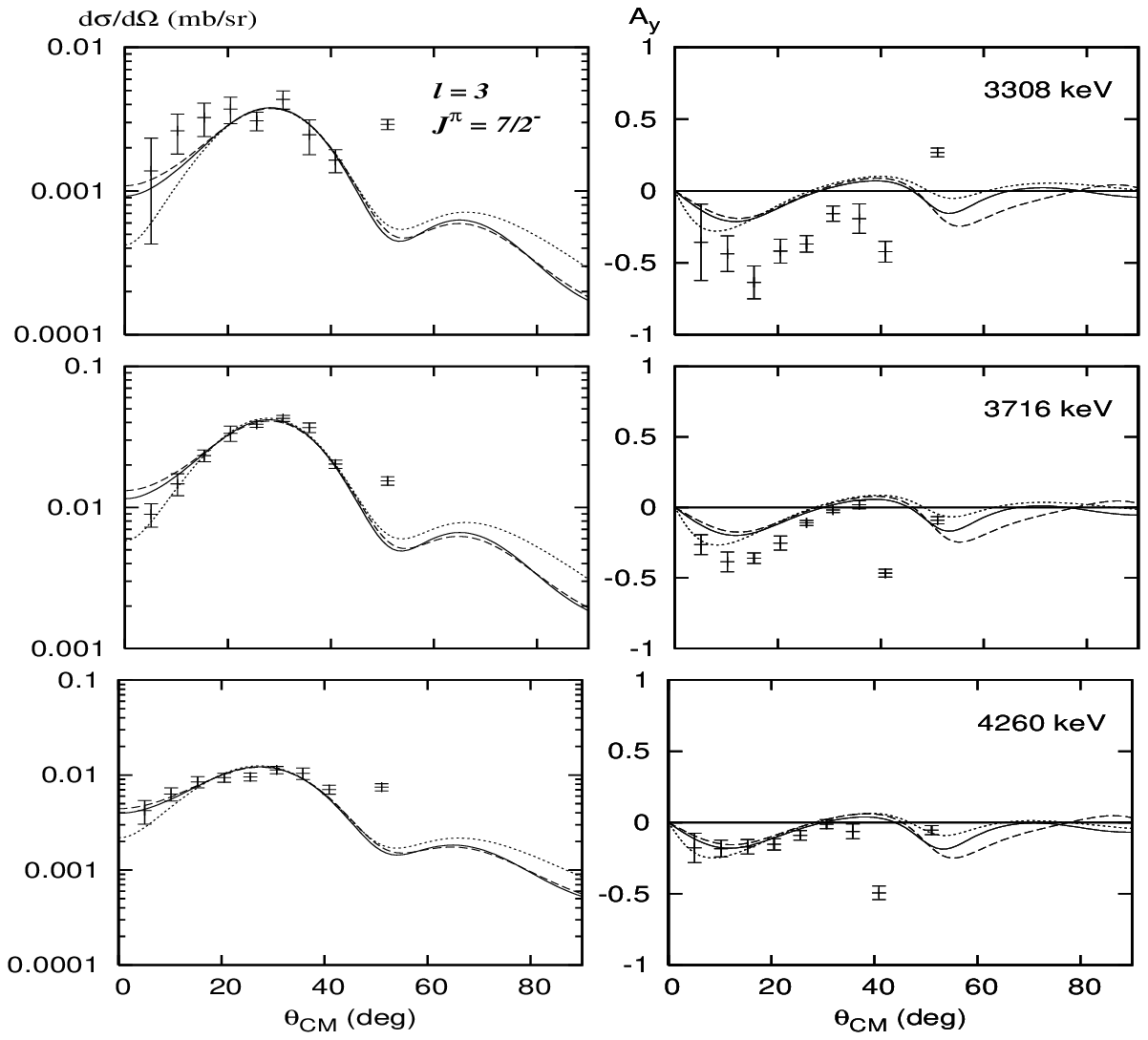}}
  \caption{Same as Fig. \ref{fig:55Fe-L1} for states with $\Delta l=3$ and  $J^\pi=7/2^-$ and energies above 3.2 MeV.
    \label{fig:55Fe-L3J7b}
    }
\end{figure}

\begin{figure}[htbp]
\resizebox{0.5\textwidth}{!}{%
 \centering
  \includegraphics{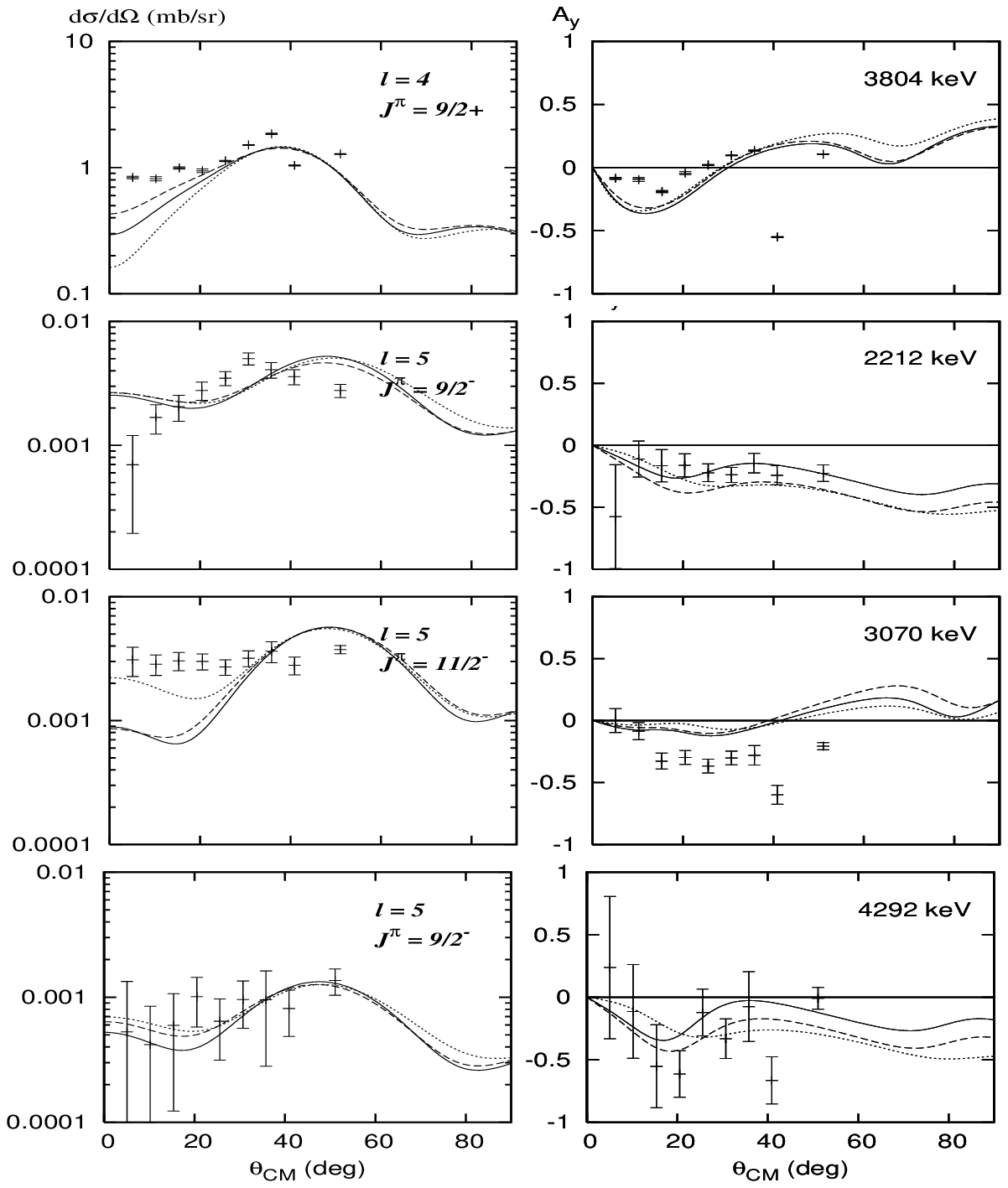}}
  \caption{Same as Fig. \ref{fig:55Fe-L1} for states with $\Delta l=4$ and $\Delta l=5$.
    \label{fig:55Fe-L4L5}
    }
\end{figure}

Angular distributions and analyzing powers are shown in figs.
\ref{fig:55Fe-L1}-\ref{fig:55Fe-L4L5}. Total angular momenta and
parities $J^{\pi}$ where deduced by comparison with results of
distorted-wave Born approximation (DWBA) calculations with the
code\\
CHUCK3\cite{Kun77}. Three different sets of optical model parameters
were used, which are summarized in table \ref{tab:OMP}.

\begin{table}[!h]
\tabcolsep 0.15cm   
\begin{center}
\caption{Optical-Model parameters for deuteron
(D1\cite{Per76},D2\cite{Tay80},D3\cite{Fes92}), proton (p)
\cite{Per76} and neutron (n) \cite{Per76} for a beam energy of
$E^{\rm lab}_{\rm d}$ = 14 MeV. V (W) denotes the real (imaginary)
potentials with volume (vol.), surface (surf.) and spin-orbit (L.S.)
terms, while r and a denote radius and diffuseness parameters.
V$_{\rm L.S.}$ was taken from Ref. \cite{Fes92}.} \label{tab:OMP}
\begin{tabular}{cccc c cccc}
\hline\noalign{\smallskip}
          &         &        & Real  &      & &        & Imaginary &       \\\cline{3-5}\cline{7-9}
          &         & V      & r$_o$ & a$_o$ & & W      & r$_I$    & a$_I$     \\
          &         & (MeV)  & (fm)  & (fm) & & (MeV)  & (fm)     & (fm)  \\ \hline
\\
D1        & vol.    & -91.68 & 1.15 & 0.81 & &        &          &       \\
          & surf.   &        &      &      & &  17.76 & 1.34     & 0.68  \\
          & L.S.     & -6.92  & 1.07 & 0.66 & &        &          &       \\
\\
D2        & vol.  & -92.64 & 1.05  & 0.86 & &        &          &       \\
          & surf. &        &       &      & &  15.26 & 1.43     & 0.69  \\
          & L.S.     & -7.00  & 0.75  & 0.5  & &        &          &       \\
\\
D3         & vol.  & -90.91 & 1.17 & 0.73 & &  -0.25 & 1.33     & 0.74  \\
          & surf. &        &      &      & &  12.32 & 1.33     & 0.79  \\
          & L.S.     & -6.92  & 0.75 & 0.5  & &        &          &       \\
\\
p         & vol.    & -53.16 & 1.17 & 0.75 & & -0.38  & 1.32     & 0.51  \\
          & surf.   &        &      &      & &  8.75  & 1.32     & 0.51  \\
          & L.S.     & -6.2   & 1.01 & 0.75 & &        &          &       \\
\\
n         & vol.    & 1.00   & 1.17 & 0.75 & &        &          &       \\
          & L.S.     & 0.00   & 1.26 & 0.69 & &        &          &       \\ \hline
\end{tabular}
\end{center}
\end{table}

The non-locality parameter $\beta$ for the particles mentioned in
tab. \ref{tab:OMP} is taken from Ref. \cite{Pol89}. The finite-range
(R) parameters for the (d,p) reaction was set to zero since it was
not used in Refs. \cite{Jun08} and \cite{Tay80}. Including it in the
calculations didn't significantly change the results.

\begin{table}[!h]
\begin{center}
\tabcolsep 0.10cm   
\caption{Transferred orbital angular momentum $\Delta l$, assigned
total angular momentum and parity $ J^{\pi}$, measured maximum
differential cross section $(\frac{\rm d \sigma}{\rm d \Omega})
^{\rm max}$ at center of mass angle $\theta^{\rm cm}_{\rm max}$ as
well as averaged absolute spectroscopic factors for the three
optical model parameter sets from Refs. \cite{Per76,Tay80,Fes92}.
For comparison literature values for $ J^{\pi}_{\rm Lit.}$
\cite{Jun08} and absolute spectroscopic factors $S_{\rm Lit.}$ are
given. The uncertainties in the spectroscopic factors reflect the
standard deviation of the three values over which was averaged. New
levels are indicated with an asterix. The uncertainties given for
spectroscopic factors do not include additional systematic
uncertainties of the cross-sections on the order of 10-20\% mainly
due to uncertainties in the target thickness.\label{tab:55fe-1}}
\begin{tabular}{ll lll |  l| l}   \hline
$E_{\rm x}^{a)} $ &
$\Delta l$  &
$ J^{\pi}$ &
$ J^{\pi}_{\rm Lit.}$ &
$(\frac{\rm d \sigma}{\rm d \Omega}) ^{\rm max}$ [$\frac{\rm mb}{\rm sr}$] &

$S_{\rm Aver.}$ &

$S_{\rm Lit.}$  \\

 [$\rm keV$] & [$\hbar$] &  [$\hbar$]    &     [$\hbar$]    & $\theta^{\rm cm}_{\rm max}$ [$^{\circ}$] &             &          \\ \hline

0.0      & 1 & $\frac{3}{2}^-$ & $\frac{3}{2}^-$ & 14.67(2)  & 0.49(1)   & 0.575$^{b)}$  \\
                &   &             &            & 15.3      &           & 0.775$^{c)}$ \\
411.0    & 1 & $\frac{1}{2}^-$ & $\frac{1}{2}^-$ & 5.53(5)   & 0.35(1)   & 0.30$^{b)}$  \\
                &   &             &            & 10.2      &           & 0.60$^{c)}$  \\
931.4    & 3 & $\frac{5}{2}^-$ & $\frac{5}{2}^-$ & 1.39(1)   & 0.43(2)   & 0.35$^{b)}$   \\
                &   &             &            & 30.6      &           & 0.65$^{c)}$   \\
1316.5   & 3 & $\frac{7}{2}^-$ & $\frac{7}{2}^-$ & 0.156(3)  & 0.028(1)  & 0.037$^{b)}$  \\
                &   &             &            & 35.7      &           & 0.045$^{c)}$    \\
1408.2   & 3 & $\frac{7}{2}^-$ & $\frac{7}{2}^-$ & 0.066(2)  & 0.012(1)  & 0.015$^{b)}$  \\
                &   &             &            & 30.6      &           & 0.018$^{c)}$   \\
1918.1   & 1 & $\frac{1}{2}^-$ & $\frac{1}{2}^-$ & 0.731(7)  & 0.0400(3) & 0.04$^{b)}$  \\
              &   &           &                & 10.2      &           & 0.10$^{c)}$  \\
2050.1   & 1 & $\frac{3}{2}^-$ & $\frac{3}{2}^-$ & 1.64(1)   & 0.046(1)  & 0.065$^{b)}$ \\
              &   &           &                & 10.2      &           & 0.088$^{c)}$ \\
2144.0   & 3 & $\frac{5}{2}^-$ & $\frac{5}{2}^-$ & 0.371(4)  & 0.101(4)  & 0.098$^{b)}$ \\
              &   &           &                & 30.7      &           & 0.153$^{c)}$ \\
2211.5   &(5)&($\frac{9}{2}^-$)& $\frac{9}{2}^-$ & 0.005(1)  & 0.0032(6) &           \\
              &   &           &                & 30.7      &           &           \\
2332.2*  & 1 & $\frac{3}{2}^-$ &                 & 0.003(1)  & 0.00007(1)&         \\
              &   &           &                & 10.2      &           &           \\
2470.2   & 1 & $\frac{3}{2}^-$ & $\frac{3}{2}^-$ & 3.65(2)   & 0.0963(2) & 0.098$^{b)}$ \\
              &   &           &                & 10.2      &           & 0.17$^{c)}$  \\
2503.8*  & 1 & $\frac{3}{2}^-$ &                 & 0.007(1)  & 0.00020(1)&          \\
              &   &           &                & 15.3      &           &           \\
2579.2   & 3 & $\frac{5}{2}^-$ & $\frac{5}{2}^-$ & 0.090(2)  & 0.0224(9) &  0.05$^{c)}$          \\
              &   &           &                & 30.7      &           &  0.03$^{d)}$  \\
2872.3   & 3 & $\frac{5}{2}^-$ & $\frac{5}{2}^-$,$\frac{7}{2}^-$& 0.040(2) & 0.0102(5) & \\
              &   &           &                & 25.6      &           &            \\
2938.9   & 3 & $\frac{7}{2}^-$ & $\frac{7}{2}^-$ & 0.195(3)  & 0.0280(5) &            \\
              &   &           &                & 30.7      &           &            \\
3028.2   & 1 & $\frac{3}{2}^-$ & $\frac{3}{2}^-$ & 0.487(7)  & 0.0119(2) &  0.021$^{d)}$         \\
              &   &           &                & 10.2      &           &           \\
3070.2   & (5)  &  ($\frac{11}{2}^-$)  & $\frac{11}{2}^-$& 0.004(1)  & 0.0031(5) &           \\
              &   &           &                & 51.0      &           &           \\
\hline

\end{tabular}
\end{center}
 {\footnotesize
 $^{a)}$ Energies from this work with uncertainties of 0.5 keV. New levels are marked with $*$.\\
 $^{b)}$ Ref.\cite{Jun08}\\
 $^{c)}$ Ref. \cite{Tay80}\\
 $^{d)}$ Ref. \cite{Ful63}\\
 }
\end{table}

Spectroscopic factors $S$ can be determined by dividing the
experimental cross-section $\sigma_{\rm (Exp.)}$ by those obtained
theoretically from DWBA calculations $\sigma_{\rm (DWBA)}$ with\\
CHUCK3:

\begin{equation}
\sigma_{\rm (Exp.)} = S * \sigma_{\rm (DWBA)}.
\label{eq:sf}
\end{equation}
In our case, $S$ was determined by a $\chi^2$ fit to the angular
distributions.

The agreement of calculated angular distributions and analyzing
powers with the experimental data is quite satisfactory for the
different sets of optical model parameters. One may argue that the
description based on the parameters of Ref.\cite{Tay80} is slightly
inferior. Tables \ref{tab:55fe-1} and \ref{tab:55fe-2} summarize the
results for all observed levels in $^{55}$Fe. The excitation
energies for those known levels that have uncertainties of less than
1 keV in Ref. \cite{Jun08} agreed for most states within less than
0.5 keV with the exceptions of the 2051.7 (4) keV, 2577.7(4) keV,
and 3072.0(4) keV states for which energies of 2050.1(5) keV,
2579.2(5) keV and 3070.2(5) keV were measured in this work,
respectively. Above 3.1 MeV we estimate the uncertainties of the
measured energies to be 2 keV.

In total, 19 previous spin assignments could be confirmed and for
six levels (2872, 3107, 3308, 3777, 3804, 4043 keV) with tentative
or several possible spin assignments a firm spin and parity
assignment could be made. For seven known levels (3354, 3655, 3716,
4019, 4117, 4134, 4372 keV) spin and parity were determined for the
first time while seven levels (2332, 2504, 3576, 3827, 3939, 4260,
4292 keV) were newly observed. It was possible to obtain
spectroscopic factors for all 39 observed levels, which are also
listed in tables \ref{tab:55fe-1} and \ref{tab:55fe-2}. The
uncertainties given for the spectroscopic factors in tables
\ref{tab:55fe-1} and \ref{tab:55fe-2} refer to the standard
deviation of the three spectroscopic factor values obtained for the
different optical model parameters. As can be easily seen, the
$S$-values for the different optical model parameters differ only
slightly. However, the cross sections may contain additional
systematic uncertainties on the order of 10-20\% mainly due to
uncertainties in the target thickness. The absolute spectroscopic
factors obtained in this work generally compare favorably with those
previously reported \cite{Jun08,Tay80}. If relative spectroscopic
factors are considered, the agreement is even better.

\begin{table}[!h]
\begin{center}
\tabcolsep 0.10cm   
\caption{ Same as Tab.\ref{tab:55fe-1} for states above 3.1 MeV.
\label{tab:55fe-2}}
\begin{tabular}{ll lll |  l| l}   \hline
$E_{\rm x}^{a)} $ &
$\Delta l$  &
$ J^{\pi}$ &
$ J^{\pi}_{\rm Lit.}$ &
$(\frac{\rm d \sigma}{\rm d \Omega}) ^{\rm max}$ [$\frac{\rm mb}{\rm sr}$] &

$S_{\rm Aver.}$ &

$S_{\rm Lit.}$  \\

 [$\rm keV$] & [$\hbar$] &  [$\hbar$]    &     [$\hbar$]    & $\theta^{\rm cm}_{\rm max}$ [$^{\circ}$] &             &          \\ \hline

3107   & 3 & $\frac{7}{2}^-$ & $\frac{5}{2}^-$,$\frac{7}{2}^-$ & 0.013(1)  & 0.0020(1)  &           \\
              &   &           &                & 30.7      &            &           \\
3308   & 3 & $\frac{7}{2}^-$ & $\frac{5}{2}^-$, $\frac{7}{2}^-$& 0.004(1) & 0.00048(2) &       \\
              &   &           &                & 30.7      &            &           \\
3354   & 3 & $\frac{5}{2}^-$ &                 & 0.009(1)  & 0.0017(1)  &           \\
              &   &           &                & 35.8      &            &           \\
3553   & 1 & $\frac{3}{2}^-$ &  $\frac{3}{2}^-$& 2.982(25) & 0.0671(1)  & 0.085$^{b)}$ \\
              &   &           &                & 15.4      &            & 0.12$^{c)}$  \\
3576*  & 1 & $\frac{3}{2}^-$ &                 & 0.028(4)  & 0.00048(2) &           \\
              &   &           &                & 15.4      &            &           \\
3591   & 1 & $\frac{1}{2}^-$ & $\frac{1}{2}^-$ & 0.016(2)  & 0.00072(2) &           \\
              &   &           &                & 10.2      &            &           \\
3655   & 3 & $\frac{5}{2}^-$ &                 & 0.005(1)  & 0.00080(5) &           \\
              &   &           &                & 35.8      &            &           \\
3716   & 3 & $\frac{7}{2}^-$ &                 & 0.043(2)  & 0.0050(1)  &           \\
              &   &           &                & 30.7      &            &           \\
3777   & 3 & $\frac{5}{2}^-$ & $\frac{3}{2}^-$,${\frac{5}{2}}^-$~$^{e)}$ & 0.025(2)  & 0.00155(4) &           \\
              &   &           &  $\frac{1}{2}^+$,${\frac{3}{2}}^+$~$^{e)}$               & 25.6      &            &           \\
              &   &           &  $\frac{1}{2}^-$~$^{f)}$                                 & 25.6      &            &           \\
3791   & 1 & $\frac{1}{2}^-$ & $\frac{1}{2}^-$ & 7.87(6)   & 0.3208(6)  &   0.5$^{c)}$        \\
              &   &           &                & 10.2      &            &  $\approx$0.7$^{g)}$  \\
3804   & 4 & $\frac{9}{2}^+$ & $\frac{7}{2}^+$,${\frac{9}{2}}^+$& 1.85(2)& 0.2734(7) &\\
              &   &           &                & 35.8      &            &           \\
3827*  & 1 & $\frac{1}{2}^-$ &                 & 0.027(4)  & 0.00117(3) &           \\
              &   &           &                & 15.4      &            &           \\
3907   & 1 & $\frac{3}{2}^-$ & $\frac{3}{2}^-$ & 0.299(11) & 0.0065(1)  &    0.018$^{g)}$       \\
              &   &           &                & 10.2      &            &           \\
3939*   & 2 &($\frac{5}{2}^+$)&                 & 0.002(1)  & 0.00004(1) &           \\
              &   &           &                & 25.6      &            &           \\
4019   & 3 & $\frac{5}{2}^-$ &                 & 0.206(4)  & 0.043(2)   &    0.06$^{c)}$       \\
              &   &           &                & 30.7      &            &   \\
4043   & 3 & $\frac{5}{2}^-$ & $\frac{5}{2}^-$,$\frac{7}{2}^-$& 0.009(1)&  0.00140(3)& 0.056$^{g)}$\\
              &   &           &                & 35.8      &            &           \\
4117   & 1 & $\frac{3}{2}^-$ &                 & 0.636(9)  & 0.0137(2)  &           \\
              &   &           &                & 15.4      &            &           \\
4134   & 3 & $\frac{5}{2}^-$ &                 & 0.033(2)  & 0.0066(3)  &           \\
              &   &           &                & 30.7      &            &           \\
4260*  & 3 & $\frac{7}{2}^-$ &                 & 0.011(1)  & 0.00130(2) &           \\
              &   &           &                & 30.7      &            &           \\
4292*  & 5 & $\frac{9}{2}^-$ &                 & 0.001(0)  & 0.0008(1)  &           \\
              &   &           &                & 51.1      &            &           \\
4372   & 2 &($\frac{5}{2}^+$)&                 & 0.009(1)  & 0.00022(1) &           \\
              &   &           &                & 15.4      &            &           \\
4450    & 2 & $\frac{5}{2}^+$ & $\frac{5}{2}^+$ & 5.80(9)   & 0.24(2)    &   0.13$^{c)}$        \\
              &   &           &                & 5.1       &            &  0.17$^{g)}$  \\

\hline

\end{tabular}
\end{center}
 {\footnotesize
 $^{a)}$ Energies from this work with uncertainties of 2 keV. New levels are marked with $*$.\\
 $^{b)}$ Ref.\cite{Jun08}\\
 $^{c)}$ Ref. \cite{Tay80}\\
 $^{d)}$ Ref. \cite{Koc72}\\
 $^{e)}$ Ref. \cite{Pet72}\\
 $^{f)}$ Ref. \cite{Max66}\\
 $^{g)}$ Ref. \cite{Ful63}\\
  }
\end{table}

Hereafter, the assignments for those levels will be discussed, for
which  the relation to known levels is not clear or where the
assignment found in this work is in conflict with previous
assignments.

{\it The 2212 keV level: } This level has a spin and parity
assignment of $J^{\pi}=\frac{9}{2}^-$ \cite{Jun08}. However, the
agreement with the DWBA calculations for $\Delta l=5$ and
$J^{\pi}=\frac{9}{2}^-$ in fig. \ref{fig:55Fe-L4L5} is not really
satisfactory. However, the measured cross section is  less than
0.005 mb and multi-step processes may be important in this case.

{\it The 3070 keV level: } There is a 3072.0(4) keV state known with
$J^{\pi}=\frac{11}{2}^-$ \cite{Jun08}, established from gamma-ray
spectroscopy \cite{Saw72,Pol74}. The measured angular distribution
is rather flat, indicating a large angular momentum transfer and the
analyzing power shows reasonable agreement with the
$J^{\pi}=\frac{11}{2}^-$ assignment. However, also in this case the
level is populated very weakly and it is not possible to exclude
multi-step processes. Therefore, a definite spin and parity
assignment can not be made on the basis of the present data.

{\it The 3354 keV $J^{\pi}=\frac{7}{2}^-$ level: } We associate this
level with the 3362(10) keV level previously observed in the (d,p)
study of Ref. \cite{Spe64}.

{\it The 3591 keV $J^{\pi}=\frac{1}{2}^-$ level: } We associate this
level with the previously observed level at 3599(10) keV
\cite{Jun08}.

{\it The 3655 keV $J^{\pi}=\frac{5}{2}^-$ level: } In Ref.
\cite{Jun08} a level is reported at 3660.8(11) keV based on the
observation of gamma rays populating this level following the
$^{51}$V($^{7}$Li,3n) reaction.  In Ref. \cite{Jun08} this level is
associated also with the 3661(10) keV level observed in (d,p)
\cite{Spe64}. No spin and parity assignment was made for this level
in either study. However, we do not observe a level at 3661 keV but
rather at 3655(2) keV. Therefore, we suggest that there are two
levels, one at 3660.8(11) keV and one at 3655(2) keV.

{\it The 3716 keV $J^{\pi}=\frac{7}{2}^-$ level: } A state at
3722(10) keV has been previously observed in (d,p) \cite{Jun08} but
no $J^{\pi}$ was assigned. We associate the 3716(2) keV level from
this work with this state.

{\it The 3777 keV $J^{\pi}=\frac{5}{2}^-$ level: } A 3770 keV level
 was reported in (p,t) experiments and assigned as
$\frac{1}{2}^-$ \cite{Max66} on the basis of observed $\Delta l=0$
angular distribution while in Ref. \cite{Pet72} a state at the same
energy with $\Delta l=1,2$ and possible spin assignments of
$(\frac{3}{2}^-,\frac{5}{2}^-)$ or $(\frac{1}{2}^+,\frac{3}{2}^+)$
was reported. From the angular distribution and analyzing power of
this level we clearly assign it as $J^{\pi}=\frac{5}{2}^-$.

{\it The 3804 keV $J^{\pi}=\frac{9}{2}^+$ level: } Ref.\cite{Jun08}
reports levels at 3814(10) keV ($\frac{7}{2}^+, \frac{9}{2}^+$), and
3815(15) ($\frac{5}{2}^-,\frac{7}{2}^-$). In earlier studies
\cite{Spe64,Koc72} a line doublet at 3.80 and 3.81 MeV was reported,
which was analyzed in Ref. \cite{Koc72} by a superposition of two
states with $J^{\pi}=\frac{1}{2}^-$ and
$J^{\pi}=(\frac{7}{2}^+,\frac{9}{2}^+)$, respectively, consistent
with our observation of the 3791(2) keV $\frac{1}{2}^-$ and 3804(2)
keV $\frac{9}{2}^+$ levels observed in this work. The 3815(10) keV
$\frac{5}{2}^-,\frac{7}{2}^-$ level listed in the Nuclear Data
Sheets \cite{Jun08} is supposedly observed in the ($^3$He,$\alpha$)
reaction. However, the original work \cite{Zam80} only reports the
3814(20) keV $\frac{9}{2}^+$ state. Therefore, one may come to the
conclusion that the 3815(10) keV $\frac{5}{2}^-,\frac{7}{2}^-$ level
should be removed from the Nuclear Data Sheets.

{\it The 4019 keV $J^{\pi}=\frac{5}{2}^-$ level: } We associate this
level with the previously observed level at 4028(10) keV
\cite{Jun08}.

{\it The 4043 keV $J^{\pi}=\frac{5}{2}^-$ level: } Ref.\cite{Jun08}
reports a level at 4057(10) keV with the energy from \cite{Spe64}
and a spectroscopic factor resulting from the averaged value of
Refs. \cite{Koc72,Ful63}. However, Ref. \cite{Koc72} reports this
$(\frac{5}{2}^-)$ level at 4.04 MeV while Ref. \cite{Ful63} reports
it an energy of 4.039 MeV, both values consistent with our
observation.

{\it The 4117 keV $J^{\pi}=\frac{3}{2}^-$ level: } We associate this
level with the previously observed level at 4110(10) keV
\cite{Jun08}.

{\it The 4134 keV $J^{\pi}=\frac{5}{2}^-$ level: } We associate this
level with the previously observed level at 4123(10) keV
\cite{Jun08}.

{\it The 4372 keV $J^{\pi}=(\frac{5}{2}^+)$ level: } We associate
this level with the previously observed level at 4372(10) keV
\cite{Jun08}. Since the agrement of the experimental analyzing power
with the DWBA results for a $\frac{5}{2}^+$ state is not very good,
we consider this assignment as tentative.

{\it The 4450 keV $J^{\pi}=\frac{5}{2}^+$ level: } We associate this
level with the previously observed  $ \frac{5}{2}^+$ level at
4463(10) keV \cite{Jun08}.

\section{Discussion}

The main aim of the current study was to investigate the role of
cross-shell excitations for the $N=28$ shell gap in $^{55}$Fe. For
this purpose we have performed large scale shell model calculations
using the code ANTOINE \cite{Cau99} and employing the GXFP1
effective interaction \cite{Hon02,Hon04} which was adjusted to the
tremendous number of experimental data available for the pf-shell
nuclei.
 The calculations were performed in the full pf-shell
({\em i.e.}, f$_{7/2}$, p$_{3/2}$, f$_{5/2}$, and p$_{1/2}$
single-particle orbitals), with up to $n=6$ particles allowed to be
excited from the f$_{7/2}$ orbital to the p$_{3/2}$, p$_{1/2}$, and
f$_{5/2}$ orbitals. Thus the 15-body wave function of $^{55}$Fe,
corresponding to the space of valence nucleons, is a superposition
of the (f$_{7/2})^{(14-k)}($p$_{3/2}$p$_{1/2}$f$_{5/2})^{1+k}$
components, where $k$ is running from 0 (no cross-shell excitation)
to $n=6$. We have found that the agreement of the calculated level
energies and spectroscopic factors with the experimental data
improves with increasing $n$ and the $n=6$ approximation (6p-6h)
yields a reasonably good description of the experimental data. the
spectroscopic factors for each $n$ were calculated using the
wave-function of the shell-model calculations for the same $n$. In
this case the ground state of $^{55}$Fe contains only $57\%$ of the
wave function of the $k=0$ configuration. Also for all excited
states the $k=0$ configuration is always less than $60\%$. It is
interesting to note that the first excited $\frac{7}{2}^-$ state
contains only $3\%$ of the $k=0$ configuration (two proton holes in
$f_{7/2}$ shell) while the second $\frac{7}{2}^-$ state contains a
considerably larger fraction ($47\%$) of the $k=0$ configuration.
The calculated spectroscopic factor for the $J^\pi=\frac{3}{2}^-$
ground state amounts to 0.87 in the $n=0$ case and reduces to 0.62
for $n=6$, which is considerably closer to the experimental value of
$S=0.49(1)$.

\begin{figure}[htbp]
\resizebox{0.5\textwidth}{!}{%
 \centering
  \includegraphics{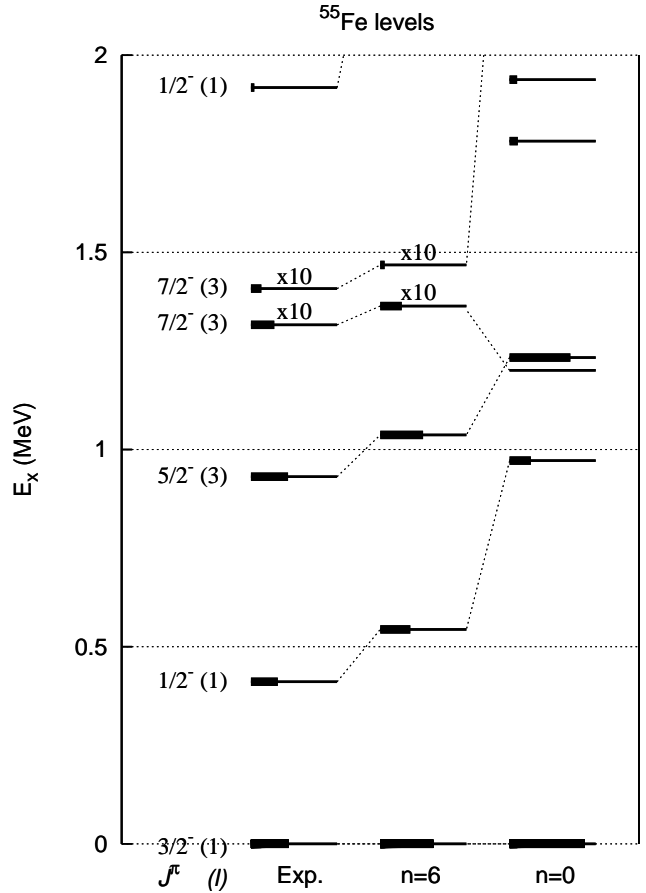}}
  \caption{$^{55}$Fe experimental levels compared to shell model predictions.
  Thick horizontal bars represent the absolute spectroscopic factors,
  with $S = 1$ corresponding to the full length of the thin level line. The
  shell model results for $n=0$ and $n=6$ cases are labeled correspondingly.
    \label{fig:55Fe-SM}
    }
\end{figure}

Fig. \ref{fig:55Fe-SM} shows the energy levels below 2 MeV from
shell model calculations for $n=0$ and $n=6$ in comparison with the
experimental results. Also shown as horizontal bars are the absolute
spectroscopic factors for the $^{54}$Fe($\overrightarrow{\rm
d}$,p)$^{55}$Fe reaction, where the full length of the thin level
line corresponds to $S=1$. For the two $\frac{7}{2}^-$ levels the
plotted spectroscopic factors are increased by a factor of 10 in
order to enhance their visibility. We note that the agreement with
the $n=6$ calculations is quite good, while the calculations based
on an assumption of an inert $N=28$ core ($n=0$) fail to reproduce
the experimental spectrum and spectroscopic factors even for the
lowest states.

\begin{figure}[htbp]
\resizebox{0.5\textwidth}{!}{%
 \centering
  \includegraphics{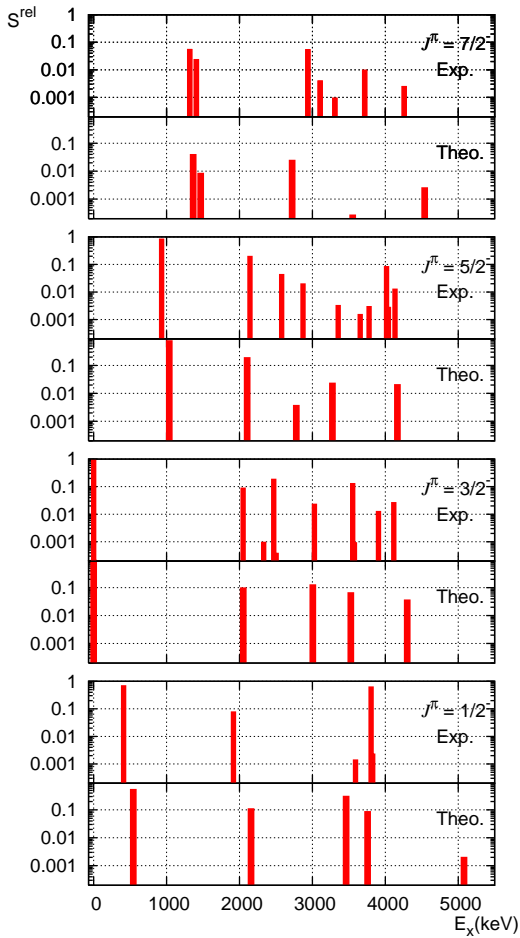}}
  \caption{Theoretical and experimental strength distribution of relative spectroscopic factors
  $S^{\rm rel.}$, normalized to the spectroscopic factor of the ground state $J^\pi={\frac{3}{2}}^-$,
   for (from bottom to top) $J^\pi={\frac{1}{2}}^-$,
   $J^\pi={\frac{3}{2}}^-$, $J^\pi={\frac{5}{2}}^-$, and $J^\pi={\frac{7}{2}}^-$ levels.
    \label{fig:55Fe-Srel}
    }
\end{figure}

The experimental and theoretical energies as well as relative
spectroscopic factors for all $J^\pi=\frac{1}{2}^-$,
$\frac{3}{2}^-$, $\frac{5}{2}^-$, and $\frac{7}{2}^-$ levels up to
an excitation energy of 5 MeV are compared in Fig.
\ref{fig:55Fe-Srel}. We note that the agreement is very good below
about 3 MeV for most spins. However, for the $J^\pi={\frac{3}{2}}^-$
states we observe already 3 levels more than we have obtained in the
shell model calculation. Furthermore, the experimental strength is
strongly fragmented already in the vicinity of 2.3 MeV. For other
spin values we observe significantly more states and stronger
fragmentation only above 3 MeV,  with the exception of the
$J^\pi={\frac{1}{2}}^-$ states. The above comparison
 clearly indicates that cross-shell excitations play an important role for the spectra and
spectroscopic properties of low-lying states in $^{55}$Fe.
Furthermore, to describe the experimentally observed density of
states below 4-5 MeV one would need to go beyond the 6p-6h
approximation adopted here and include higher np-nh excitations.

\section{Summary}
In a  high resolution study of the reaction  $^{54}{\rm
Fe}(\overrightarrow{\rm d},{\rm p})^{55}{\rm Fe}$ 39 levels were
observed. On the basis of angular distributions and asymmetries firm
spin and parity assignments could be made for all but 2 states, for
which firm assignments were already available from previous studies.
It was possible to observe eight new states and prove that five
states previously assigned to $^{55}$Fe were actually states in
$^{57}$Fe. For a number of known states energies could be determined
with improved accuracy and spins were assigned for the first time.
Spectroscopic factors were determined for all observed states. The
results were compared with large scale shell model calculations
using the GXPF1 effective interaction and good agreement was found
up to 2 MeV when up to 6p-6h excitations across the $N=28$ shell
were included, clearly establishing the significant role of
cross-shell excitations in $^{54}$Fe and $^{55}$Fe. The observed
larger number of levels and stronger fragmentation of the
spectroscopic strength above 2.5 MeV in comparison to shell model
results indicate that more many-body cross-shell degrees of freedom
need to be taken into account to describe the experimental data
obtained in this work.

\section{Acknowledgement}

The excellent work of the MLL accelerator crew is gratefully
acknowledged. Useful discussions with J. Tostevin are acknowledged.
This work was supported by DFG under grants KR2326/1-1, 436 RUM
17/1/07 and by the DFG cluster of excellence Origin and Structure of
the Universe (http://www.universe-cluster.de). A.F.L. acknowledges
partial support of this work from NSF grant PHY0555396.

\newpage


\end{document}